\documentclass[12pt]{article}
\begin{document}

\author{C. Bizdadea\thanks{%
e-mail address: bizdadea@central.ucv.ro}, E. M. Cioroianu\thanks{%
e-mail address: manache@central.ucv.ro}, S. O. Saliu\thanks{%
e-mail address: osaliu@central.ucv.ro} \\
Faculty of Physics, University of Craiova\\
13 A. I. Cuza Str., Craiova RO-1100, Romania}
\title{Hamiltonian cohomological derivation of four-dimensional nonlinear gauge
theories }
\maketitle

\begin{abstract}
Consistent couplings among a set of scalar fields, two types of one-forms
and a system of two-forms are investigated in the light of the Hamiltonian
BRST cohomology, giving a four-dimensional nonlinear gauge theory. The
emerging interactions deform the first-class constraints, the Hamiltonian
gauge algebra, as well as the reducibility relations.

\end{abstract}

\section{Introduction}

The Hamiltonian version of the BRST symmetry \cite{2}, \cite{8} imposed
itself as an appropriate setting for analysing various topics in gauge
theories, such as its implementation in quantum mechanics \cite{2} (Chapter
14), or the appropriate correlation between the BRST symmetry itself and
canonical quantization methods \cite{9}. Meanwhile, the cohomological
development of the BRST method was proved to be a useful tool for
approaching some less known aspects, like the Hamiltonian analysis of
anomalies \cite{10}, the precise relation between the local Lagrangian and
Hamiltonian BRST cohomologies \cite{11}, and, recently, the problem of
obtaining consistent Hamiltonian interactions in gauge theories by means of
the deformation theory \cite{12}.

In this paper we investigate the consistent Hamiltonian interactions that
can be added among a set of scalar fields, two types of one-forms and a
system of two-forms in four dimensions, described in the free limit by an
abelian BF theory \cite{13}, which result in a four-dimensional nonlinear
gauge theory. This paper extends our results related to the two-dimensional
case \cite{mpla}. It is known that nonlinear gauge theories in two
dimensions \cite{14} are already important as they are related to pure
two-dimensional gravitation theory \cite{15}. Indeed, when the nonlinear
algebra is the Lorentz-covariant extension of the Poincar\'{e} algebra, one
recovers nothing but the Yang-Mills-like formulation of $\mathbf{R}^{2}$
gravity with dynamical torsion, the so-called `dilaton' gravity \cite{16}.
In this light, it appears quite clear that the derivation of nonlinear gauge
theories in dimensions higher than two might bring significant contribution
to the evolvement of a conceptual mechanism for the study of quantum gravity
in higher dimensions from the perspective of gauge theories.

Our strategy is as follows. Initially, we derive the Hamiltonian BRST
symmetry of the abelian BF theory in four dimensions, which splits as the
sum between the Koszul-Tate differential and the exterior derivative along
the gauge orbits. Next, we solve the main equations governing the
Hamiltonian deformation procedure on behalf of the BRST cohomology of the
free theory. As a consequence, we find the BRST charge and BRST-invariant
Hamiltonian of the deformed model. With the help of these deformed
quantities, we identify the interacting gauge theory by analysing the
resulting first-class constraints, first-class Hamiltonian and gauge algebra.

Our paper is organized in seven sections. Section 2 briefly reviews the
problem of constructing consistent Hamiltonian interactions in the framework
of the BRST formalism. In Section 3 we derive the BRST symmetry of the free
model. In Sections 4 and 5 we compute the deformed BRST charge,
respectively, the deformed BRST-invariant Hamiltonian. Section 6 focuses on
the identification of the interacting model. Section 7 ends the paper with
the main conclusions.

\section{Main equations of the Hamiltonian deformation procedure}

It has been shown in \cite{12} that the problem of constructing consistent
Hamiltonian interactions in theories with first-class constraints can be
reformulated as a deformation problem of the BRST charge $\Omega _{0}$ and
of the BRST-invariant Hamiltonian $H_{0\mathrm{B}}$ of a given ``free''
first-class theory. If the interactions can be consistently constructed,
then the ``free'' BRST charge can be deformed as 
\begin{eqnarray}
&&\Omega _{0}\rightarrow \Omega =\Omega _{0}+g\int d^{D-1}x\omega
_{1}+g^{2}\int d^{D-1}x\omega _{2}+O\left( g^{3}\right)   \nonumber \\
&=&\Omega _{0}+g\Omega _{1}+g^{2}\Omega _{2}+O\left( g^{3}\right) ,
\label{xx39}
\end{eqnarray}
where $\Omega $ should satisfy the equation 
\begin{equation}
\left[ \Omega ,\Omega \right] =0.  \label{xx40}
\end{equation}
Equation (\ref{xx40}) splits accordingly the deformation parameter $g$ as 
\begin{equation}
\left[ \Omega _{0},\Omega _{0}\right] =0,  \label{xx41}
\end{equation}
\begin{equation}
2\left[ \Omega _{0},\Omega _{1}\right] =0,  \label{xx42}
\end{equation}
\begin{equation}
2\left[ \Omega _{0},\Omega _{2}\right] +\left[ \Omega _{1},\Omega
_{1}\right] =0,  \label{xx43}
\end{equation}
\[
\vdots 
\]
Obviously, equation (\ref{xx41}) is automatically satisfied. From the
remaining equations we deduce the pieces $\left( \Omega _{k}\right) _{k>0}$
on account of the ``free'' BRST differential. With the deformed BRST charge
at hand, we then deform the BRST-invariant Hamiltonian of the ``free''
theory 
\begin{eqnarray}
&&H_{0\mathrm{B}}\rightarrow H_{\mathrm{B}}=H_{0\mathrm{B}}+g\int
d^{D-1}xh_{1}+g^{2}\int d^{D-1}xh_{2}+O\left( g^{3}\right)   \nonumber \\
&=&H_{0\mathrm{B}}+gH_{1}+g^{2}H_{2}+O\left( g^{3}\right) ,  \label{xx44}
\end{eqnarray}
and require that 
\begin{equation}
\left[ H_{\mathrm{B}},\Omega \right] =0.  \label{xx45}
\end{equation}
Like in the previous case, equation (\ref{xx45}) can be decomposed
accordingly the deformation parameter like 
\begin{equation}
\left[ H_{0\mathrm{B}},\Omega _{0}\right] =0,  \label{xx46}
\end{equation}
\begin{equation}
\left[ H_{0\mathrm{B}},\Omega _{1}\right] +\left[ H_{1},\Omega _{0}\right]
=0,  \label{xx47}
\end{equation}
\begin{equation}
\left[ H_{0\mathrm{B}},\Omega _{2}\right] +\left[ H_{1},\Omega _{1}\right]
+\left[ H_{2},\Omega _{0}\right] =0,  \label{xx48}
\end{equation}
\[
\vdots 
\]
Clearly, equation (\ref{xx46}) is again fulfilled, while from the other
equations one can determine the components $\left( H_{k}\right) _{k>0}$ by
relying on the BRST symmetry of the ``free'' model. Once the deformations
are computed, special attention should be paid to the elimination of
non-locality, as well as of triviality of the resulting deformations.

\section{BRST symmetry of the free theory}

We begin with a free model that describes an abelian four-dimensional BF
theory involving a set of scalar fields, two types of one-forms and a
collection of two-forms 
\begin{equation}
S_{0}[A_{\mu }^{a},H_{\mu }^{a},\varphi _{a},B_{a}^{\mu \nu }]=\int
d^{4}x\left( H_{\mu }^{a}\partial ^{\mu }\varphi _{a}+\frac{1}{2}B_{a}^{\mu
\nu }\partial _{[\mu }A_{\nu ]}^{a}\right) ,  \label{xx1}
\end{equation}
where the notation $\left[ \mu \nu \right] $ signifies antisymmetry with
respect to the indices between brackets. The above action is invariant under
the gauge transformations 
\begin{equation}
\delta _{\epsilon }A_{\mu }^{a}=\partial _{\mu }\epsilon ^{a},\;\delta
_{\epsilon }H_{\mu }^{a}=\partial ^{\nu }\epsilon _{\mu \nu }^{a},\;\delta
_{\epsilon }\varphi _{a}=0,\;\delta _{\epsilon }B_{a}^{\mu \nu }=\partial
_{\rho }\epsilon _{a}^{\mu \nu \rho },  \label{xxgauge}
\end{equation}
which are off-shell second-stage reducible, where the gauge parameters $%
\epsilon ^{a}$, $\epsilon _{\mu \nu }^{a}$ and $\epsilon _{a}^{\mu \nu \rho
} $ are bosonic, the last two sets being completely antisymmetric.

After the elimination of the second-class constraints (the co-ordinates of
the reduced phase-space are $z^{A}=\left( \pi _{a}^{0},A_{\mu
}^{a},B_{a}^{\mu \nu },p_{a}^{i},H_{\mu }^{a},\pi _{ij}^{a},\varphi
_{a}\right) $), we are left with a system subject only to the first-class
constraints 
\begin{equation}
G_{a}^{(1)}\equiv \pi _{a}^{0}\approx 0,\;G_{a}^{(2)}\equiv -\partial
_{i}B_{a}^{0i}\approx 0,  \label{xx2}
\end{equation}
\begin{equation}
G_{ij}^{(1)a}\equiv \pi _{ij}^{a}\approx 0,\;G_{ij}^{(2)a}\equiv -\frac{1}{2}%
\partial _{[i}A_{j]}^{a}\approx 0,  \label{xx3}
\end{equation}
\begin{equation}
\gamma _{a}^{(1)i}\equiv p_{a}^{i}\approx 0,\;\gamma _{a}^{(2)i}\equiv
-\partial ^{i}\varphi _{a}\approx 0,  \label{xx4}
\end{equation}
and displaying the first-class Hamiltonian 
\begin{equation}
H_{0}=\int d^{3}x\left( H_{i}^{a}\gamma
_{a}^{(2)i}+B_{a}^{ij}G_{ij}^{(2)a}+A_{0}^{a}G_{a}^{(2)}\right) ,
\label{xx5}
\end{equation}
in terms of the non-vanishing fundamental Dirac brackets 
\begin{equation}
\left[ \pi _{a}^{0}(t,\mathbf{x}),A_{0}^{b}(t,\mathbf{y})\right] =-\delta
_{a}^{b}\delta ^{3}\left( \mathbf{x}-\mathbf{y}\right) ,  \label{xx6}
\end{equation}
\begin{equation}
\left[ B_{a}^{0i}(t,\mathbf{x}),A_{j}^{b}(t,\mathbf{y})\right] =-\delta
_{j}^{i}\delta _{a}^{b}\delta ^{3}\left( \mathbf{x}-\mathbf{y}\right) ,
\label{xx7}
\end{equation}
\begin{equation}
\left[ H_{0}^{a}(t,\mathbf{x}),\varphi _{b}(t,\mathbf{y})\right] =-\delta
_{b}^{a}\delta ^{3}\left( \mathbf{x}-\mathbf{y}\right) ,  \label{xx8}
\end{equation}
\begin{equation}
\left[ \pi _{ij}^{a}(t,\mathbf{x}),B_{b}^{kl}(t,\mathbf{y})\right] =-\frac{1%
}{2}\delta _{b}^{a}\delta _{i}^{[k}\delta _{j}^{l]}\delta ^{3}\left( \mathbf{%
x}-\mathbf{y}\right) ,  \label{xx9}
\end{equation}
\begin{equation}
\left[ p_{a}^{i}(t,\mathbf{x}),H_{j}^{b}(t,\mathbf{y})\right] =-\delta
_{j}^{i}\delta _{a}^{b}\delta ^{3}\left( \mathbf{x}-\mathbf{y}\right) .
\label{xx10}
\end{equation}
The above constraints are abelian, while the remaining gauge algebra
relations are expressed by 
\begin{equation}
\left[ H_{0},G_{a}^{(1)}\right] =G_{a}^{(2)},\;\left[
H_{0},G_{a}^{(2)}\right] =0,  \label{xx11}
\end{equation}
\begin{equation}
\left[ H_{0},G_{ij}^{(1)a}\right] =G_{ij}^{(2)a},\;\left[
H_{0},G_{ij}^{(2)a}\right] =0,  \label{xx12}
\end{equation}
\begin{equation}
\left[ H_{0},\gamma _{a}^{(1)i}\right] =\gamma _{a}^{(2)i},\;\left[
H_{0},\gamma _{a}^{(2)i}\right] =0.  \label{xx13}
\end{equation}
The constraint functions $G_{ij}^{(2)a}$ are first-stage reducible, with the
reducibility functions given by 
\begin{equation}
Z_{aklm}^{bij}=\frac{1}{2}\delta _{a}^{b}\partial _{[k}\delta _{l}^{i}\delta
_{m]}^{j},  \label{xx14}
\end{equation}
while the constraint functions $\gamma _{a}^{(2)i}$ are second-stage
reducible, where the first-, respectively, second-stage reducibility
functions read as 
\begin{equation}
Z_{bi}^{ajk}=\delta _{b}^{a}\partial ^{[j}\delta _{i}^{k]},\;Z_{cjk}^{blmn}=%
\frac{1}{2}\delta _{c}^{b}\partial ^{[l}\delta _{j}^{m}\delta _{k}^{n]}.
\label{xx15}
\end{equation}
We mention that all the reducibility relations hold off-shell.

The Hamiltonian BRST formalism requires the introduction of the ghosts 
\begin{equation}
\eta ^{a_{0}}=\left( \eta ^{(1)a},\eta ^{(2)a},\eta _{a}^{(1)ij},\eta
_{a}^{(2)ij},C_{i}^{(1)a},C_{i}^{(2)a}\right) ,  \label{xx16}
\end{equation}
\begin{equation}
\eta ^{a_{1}}=\left( C_{ij}^{a},\eta _{a}^{ijk}\right) ,\;\eta
^{a_{2}}=\left( C_{ijk}^{a}\right) ,  \label{xx17}
\end{equation}
together with their conjugated antighosts 
\begin{equation}
\mathcal{P}_{a_{0}}=\left( \mathcal{P}_{a}^{(1)},\mathcal{P}_{a}^{(2)},%
\mathcal{P}_{ij}^{(1)a},\mathcal{P}_{ij}^{(2)a},P_{a}^{(1)i},P_{a}^{(2)i}%
\right) ,  \label{xx18}
\end{equation}
\begin{equation}
\mathcal{P}_{a_{1}}=\left( P_{a}^{ij},\mathcal{P}_{ijk}^{a}\right) ,\;%
\mathcal{P}_{a_{2}}=\left( P_{a}^{ijk}\right) .  \label{xx19}
\end{equation}
In (\ref{xx16}--\ref{xx17}), the fields $\eta ^{a_{0}}$ and $\eta ^{a_{2}}$
are fermionic, with the ghost number equal to one, respectively, to three,
while $\eta ^{a_{1}}$ are bosonic, of ghost number two. The ghost number is
defined in the usual manner as the difference between the pure ghost number (%
$\mathrm{pgh}$) and the antighost number ($\mathrm{antigh}$), where 
\begin{equation}
\mathrm{pgh}\left( z^{A}\right) =0,\;\mathrm{pgh}\left( \eta ^{a_{0}}\right)
=1,\;\mathrm{pgh}\left( \mathcal{P}_{a_{0}}\right) =0,  \label{xx20}
\end{equation}
\begin{equation}
\mathrm{antigh}\left( z^{A}\right) =0,\;\mathrm{antigh}\left( \eta
^{a_{0}}\right) =0,\;\mathrm{antigh}\left( \mathcal{P}_{a_{0}}\right) =1,
\label{xx21}
\end{equation}
\begin{equation}
\mathrm{pgh}\left( \eta ^{a_{1}}\right) =2,\;\mathrm{pgh}\left( \mathcal{P}%
_{a_{1}}\right) =0,  \label{xx22}
\end{equation}
\begin{equation}
\mathrm{antigh}\left( \eta ^{a_{1}}\right) =0,\;\mathrm{antigh}\left( 
\mathcal{P}_{a_{1}}\right) =2,  \label{xx23}
\end{equation}
\begin{equation}
\mathrm{pgh}\left( \eta ^{a_{2}}\right) =3,\;\mathrm{pgh}\left( \mathcal{P}%
_{a_{2}}\right) =0,  \label{xx24}
\end{equation}
\begin{equation}
\mathrm{antigh}\left( \eta ^{a_{2}}\right) =0,\;\mathrm{antigh}\left( 
\mathcal{P}_{a_{2}}\right) =3.  \label{xx25}
\end{equation}
The BRST charge of the free model under discussion takes the form 
\begin{eqnarray}
&&\Omega _{0}=\int d^{3}x\left( \eta ^{(1)a}G_{a}^{(1)}+\eta
^{(2)a}G_{a}^{(2)}+\eta _{a}^{(1)ij}G_{ij}^{(1)a}+\eta
_{a}^{(2)ij}G_{ij}^{(2)a}+C_{i}^{(1)a}\gamma _{a}^{(1)i}\right.  \nonumber \\
&&\left. +C_{i}^{(2)a}\gamma _{a}^{(2)i}+C_{ij}^{a}\partial
^{[i}P_{a}^{(2)j]}+\eta _{a}^{ijk}\partial _{[i}\mathcal{P}%
_{jk]}^{(2)a}+C_{ijk}^{a}\partial ^{[i}P_{a}^{jk]}\right) ,  \label{xx26}
\end{eqnarray}
while the corresponding BRST-invariant Hamiltonian is 
\begin{equation}
H_{0\mathrm{B}}=H_{0}+\int d^{3}x\left( \eta ^{(1)a}\mathcal{P}%
_{a}^{(2)}+C_{i}^{(1)a}P_{a}^{(2)i}+\eta _{a}^{(1)ij}\mathcal{P}%
_{ij}^{(2)a}\right) .  \label{xx27}
\end{equation}
The BRST symmetry of the free theory, $s\bullet =[\bullet ,\Omega _{0}]$,
splits as 
\begin{equation}
s=\delta +\gamma ,  \label{xx28}
\end{equation}
where $\delta $ denotes the Koszul-Tate differential ($\mathrm{antigh}\left(
\delta \right) =-1$, $\mathrm{pgh}\left( \delta \right) =0$), and $\gamma $
represents the exterior longitudinal derivative ($\mathrm{antigh}\left(
\gamma \right) =0$, $\mathrm{pgh}\left( \gamma \right) =1$). These two
operators act on the variables from the BRST complex like 
\begin{equation}
\delta z^{A}=0,\;\delta \eta ^{a_{0}}=0,\;\delta \eta ^{a_{1}}=0,\;\delta
\eta ^{a_{2}}=0,  \label{xx29}
\end{equation}
\begin{equation}
\delta \mathcal{P}_{a}^{(1)}=-\pi _{a}^{0},\;\delta \mathcal{P}%
_{a}^{(2)}=\partial _{i}B_{a}^{0i},\;\delta P_{a}^{(1)i}=-p_{a}^{i},\;\delta
P_{a}^{(2)i}=\partial ^{i}\varphi _{a},  \label{xx30}
\end{equation}
\begin{equation}
\delta \mathcal{P}_{ij}^{(1)a}=-\pi _{ij}^{a},\;\delta \mathcal{P}%
_{ij}^{(2)a}=\frac{1}{2}\partial _{[i}A_{j]}^{a},\;\delta
P_{a}^{ij}=-\partial ^{[i}P_{a}^{(2)j]},  \label{xx31}
\end{equation}
\begin{equation}
\delta \mathcal{P}_{ijk}^{a}=-\partial _{[i}\mathcal{P}_{jk]}^{(2)a},\;%
\delta P_{a}^{ijk}=-\partial ^{[i}P_{a}^{jk]},  \label{xx32}
\end{equation}
\begin{equation}
\gamma A_{i}^{a}=\partial _{i}\eta ^{(2)a},\;\gamma A_{0}^{a}=\eta
^{(1)a},\;\gamma \varphi _{a}=0,\;\gamma \pi _{a}^{0}=0,\;\gamma
p_{a}^{i}=0,\;\gamma \pi _{ij}^{a}=0,  \label{xx34}
\end{equation}
\begin{equation}
\gamma B_{a}^{0i}=\partial _{j}\eta _{a}^{(2)ij},\;\gamma B_{a}^{ij}=\eta
_{a}^{(1)ij},\;\gamma H_{i}^{a}=C_{i}^{(1)a},\;\gamma H_{0}^{a}=-\partial
^{i}C_{i}^{(2)a},  \label{xx35}
\end{equation}
\begin{equation}
\gamma \eta ^{(1)a}=\gamma \eta ^{(2)a}=\gamma C_{i}^{(1)a}=\gamma \eta
_{a}^{(1)ij}=0,  \label{xx36}
\end{equation}
\begin{equation}
\gamma C_{i}^{(2)a}=-2\partial ^{j}C_{ij}^{a},\;\gamma C_{ij}^{a}=-3\partial
^{k}C_{ijk}^{a},\;\gamma \eta _{a}^{(2)ij}=3\partial _{k}\eta _{a}^{ijk},
\label{xx37}
\end{equation}
\begin{equation}
\gamma \eta _{a}^{ijk}=0,\;\gamma C_{ijk}^{a}=0,\;\gamma \mathcal{P}%
_{a_{0}}=0,\;\gamma \mathcal{P}_{a_{1}}=0,\;\gamma \mathcal{P}_{a_{2}}=0.
\label{xx38}
\end{equation}
The last formulas will be employed in the next section for the deformation
of the free theory.

\section{Deformed BRST charge}

In this section we approach the equations that govern the deformation of the
BRST charge by relying on cohomological techniques. As a result, we find
that only the first-order deformation is non-trivial, while its consistency
reveals the Jacobi identity for a nonlinear algebra.

\subsection{First-order deformation}

Initially, we solve the equation (\ref{xx42}), which is responsible for the
first-order deformation of the BRST charge. It takes the local form 
\begin{equation}
s\omega _{1}=\partial _{i}j^{i},  \label{xx49}
\end{equation}
for some local $j^{i}$. In order to investigate this equation, we develop $%
\omega _{1}$ according to the antighost number 
\begin{equation}
\omega _{1}=\stackrel{(0)}{\omega }_{1}+\stackrel{(1)}{\omega }_{1}+\cdots +%
\stackrel{(J)}{\omega }_{1},\;\mathrm{antigh}\left( \stackrel{(I)}{\omega }%
_{1}\right) =I,  \label{xx50}
\end{equation}
where the last term in the last formula can be assumed to be annihilated by $%
\gamma $, $\gamma \stackrel{(J)}{\omega }_{1}=0$. Thus, we need to know the
cohomology of $\gamma $, $H\left( \gamma \right) $, for computing the piece
of highest antighost number in (\ref{xx50}). From the definitions of $\gamma 
$ acting on the generators of the BRST complex (see the relations (\ref{xx34}%
--\ref{xx38})), we remark that $H\left( \gamma \right) $ is generated by 
\begin{equation}
\Phi ^{\alpha }=\left( F_{ij}^{a}=\partial _{\left[ i\right. }A_{\left.
j\right] }^{a},\varphi _{a},\pi _{a}^{0},p_{a}^{i},\pi _{ij}^{a},\partial
_{i}B_{a}^{0i}\right) ,  \label{xx51}
\end{equation}
together with their spatial derivatives, by the antighosts (\ref{xx18}--\ref
{xx19}) and their derivatives, as well as by the undifferentiated ghosts $%
\eta ^{(2)a}$, $\eta _{a}^{ijk}$ and $C_{ijk}^{a}$. Consequently, the
general solution to the equation $\gamma a=0$ can be written (up to a
trivial term) as 
\begin{equation}
a=a_{M}\left( \left[ \Phi ^{\alpha }\right] ,\left[ \mathcal{P}%
_{a_{0}}\right] ,\left[ \mathcal{P}_{a_{1}}\right] ,\left[ \mathcal{P}%
_{a_{2}}\right] \right) e^{M}\left( \eta ^{(2)a},\eta
_{a}^{ijk},C_{ijk}^{a}\right) ,  \label{xx52}
\end{equation}
where $e^{M}\left( \eta ^{(2)a},\eta _{a}^{ijk},C_{ijk}^{a}\right) $ stands
for a basis in the space of the polynomials in the ghosts. The notation $%
f\left[ q\right] $ signifies that $f$ depends on $q$ and its spatial
derivatives up to a finite order. As $\mathrm{antigh}\left( \stackrel{(J)}{%
\omega }_{1}\right) =J$ and $\mathrm{gh}\left( \stackrel{(J)}{\omega }%
_{1}\right) =1$, it follows that $\mathrm{pgh}\left( \stackrel{(J)}{\omega }%
_{1}\right) =J+1$. Thus, using (\ref{xx52}), it results that the general
solution to the equation $\gamma \stackrel{(J)}{\omega }_{1}=0$ is 
\begin{equation}
\stackrel{(J)}{\omega }_{1}=a_{J}\left( \left[ \Phi ^{\alpha }\right]
,\left[ \mathcal{P}_{a_{0}}\right] ,\left[ \mathcal{P}_{a_{1}}\right]
,\left[ \mathcal{P}_{a_{2}}\right] \right) e^{J+1}\left( \eta ^{(2)a},\eta
_{a}^{ijk},C_{ijk}^{a}\right) ,  \label{xx53}
\end{equation}
where $\mathrm{antigh}\left( a_{J}\right) =J$.

The equation (\ref{xx49}) projected on antighost number $\left( J-1\right) $
reads as 
\begin{equation}
\delta \stackrel{(J)}{\omega }_{1}+\gamma \stackrel{(J-1)}{\omega }%
_{1}=\partial _{i}n^{i}.  \label{xx54}
\end{equation}
For the equation (\ref{xx54}) to possess solutions (in other words, for $%
\stackrel{(J-1)}{\omega }_{1}$ to exist), it is necessary that the functions 
$a_{J}$ belong to $H_{J}\left( \delta |\tilde{d}\right) $, where $%
H_{J}\left( \delta |\tilde{d}\right) $ means the homological space
containing objects of antighost number equal to $J$ that are $\delta $%
-closed modulo the spatial part of the exterior space-time derivative $%
\tilde{d}$. Translating the Lagrangian results from \cite{gen} at the
Hamiltonian level, we have that 
\begin{equation}
H_{J}\left( \delta |\tilde{d}\right) =0,\;\mathrm{for}\;J>3,  \label{xx55}
\end{equation}
so we can assume that the development (\ref{xx50}) stops after the first
four terms 
\begin{equation}
\omega _{1}=\stackrel{(0)}{\omega }_{1}+\stackrel{(1)}{\omega }_{1}+%
\stackrel{(2)}{\omega }_{1}+\stackrel{(3)}{\omega }_{1},  \label{xx56}
\end{equation}
where $\stackrel{(3)}{\omega }_{1}$ results from (\ref{xx53}) with $J=3$. In
the meantime, the most general representative of $H_{3}\left( \delta |\tilde{%
d}\right) $ is given by 
\begin{equation}
a_{3}=k_{ijk}\left( \frac{\delta U}{\delta \varphi _{a}}P_{a}^{ijk}-\frac{%
\delta ^{2}U}{\delta \varphi _{a}\delta \varphi _{b}}P_{a}^{[i}P_{b}^{jk]}-%
\frac{\delta ^{3}U}{\delta \varphi _{a}\delta \varphi _{b}\delta \varphi _{c}%
}P_{a}^{i}P_{b}^{j}P_{c}^{k}\right) ,  \label{xx57}
\end{equation}
with $k_{ijk}$ some antisymmetric constants, and $U$ an arbitrary function
that depends on $\varphi _{a}$, but not on their derivatives. Now, we can
completely determine the last component in (\ref{xx56}). On the one hand,
the elements of $e^{4}\left( \eta ^{(2)a},\eta _{a}^{ijk},C_{ijk}^{a}\right) 
$ can be written in the form 
\begin{equation}
C_{ijk}^{a}\eta ^{(2)b},\eta _{a}^{ijk}\eta _{b}^{i^{\prime }j^{\prime
}k^{\prime }},\eta _{a}^{ijk}\eta ^{(2)b}\eta ^{(2)c},\eta ^{(2)a}\eta
^{(2)b}\eta ^{(2)c}\eta ^{(2)d}.  \label{xx58}
\end{equation}
On the other hand, we ask that the resulting deformations are covariant and
independent of the space-time dimension. In view of this, the second and
fourth elements in (\ref{xx58}) should be discarded as they need a
three-dimensional antisymmetric symbol in order to be glued to (\ref{xx57}).
Then, we obtain that 
\begin{eqnarray}
&&\stackrel{(3)}{\omega }_{1}=-\frac{1}{4}\left( \frac{\delta M_{bc}^{a}}{%
\delta \varphi _{d}}P_{d}^{ijk}-\frac{\delta ^{2}M_{bc}^{a}}{\delta \varphi
_{d}\delta \varphi _{e}}P_{d}^{(2)[i}P_{e}^{jk]}\right.   \nonumber \\
&&\left. -\frac{\delta ^{3}M_{bc}^{a}}{\delta \varphi _{d}\delta \varphi
_{e}\delta \varphi _{f}}P_{d}^{(2)i}P_{e}^{(2)j}P_{f}^{(2)k}\right) \eta
_{aijk}\eta ^{(2)b}\eta ^{(2)c}-\left( \frac{\delta W_{ab}}{\delta \varphi
_{c}}P_{c}^{ijk}\right.   \nonumber \\
&&\left. -\frac{\delta ^{2}W_{ab}}{\delta \varphi _{c}\delta \varphi _{d}}%
P_{c}^{(2)[i}P_{d}^{jk]}-\frac{\delta ^{3}W_{ab}}{\delta \varphi _{c}\delta
\varphi _{d}\delta \varphi _{e}}P_{c}^{(2)i}P_{d}^{(2)j}P_{e}^{(2)k}\right)
C_{ijk}^{a}\eta ^{(2)b},  \label{xx59}
\end{eqnarray}
where $M_{bc}^{a}$ and $W_{ab}$ depend on $\varphi _{a}$, with $M_{bc}^{a}$
antisymmetric in its lower indices. With $\stackrel{(3)}{\omega }_{1}$ at
hand, we pass to determining the piece of antighost number two from the
first-order deformation. It obeys the equation (\ref{xx54}) with $J=3$, that
leads to 
\begin{eqnarray}
&&\stackrel{(2)}{\omega }_{1}=\left( \frac{\delta W_{ab}}{\delta \varphi _{c}%
}P_{c}^{ij}+\frac{\delta ^{2}W_{ab}}{\delta \varphi _{c}\delta \varphi _{d}}%
P_{c}^{(2)i}P_{d}^{(2)j}\right) \left( C_{ij}^{a}\eta
^{(2)b}+3C_{ijk}^{a}A^{bk}\right)   \nonumber \\
&&+\frac{1}{2}\left( \frac{\delta M_{bc}^{a}}{\delta \varphi _{d}}P_{d}^{ij}+%
\frac{\delta ^{2}M_{bc}^{a}}{\delta \varphi _{d}\delta \varphi _{e}}%
P_{d}^{(2)i}P_{e}^{(2)j}\right) \left( \frac{1}{2}\eta _{aij}^{(2)}\eta
^{(2)b}\eta ^{(2)c}\right.   \nonumber \\
&&\left. +3\eta _{aijk}\eta ^{(2)b}A^{ck}\right) +\left( M_{bc}^{a}\mathcal{P%
}_{ijk}^{c}+\frac{\delta M_{bc}^{a}}{\delta \varphi _{d}}P_{d[i}^{(2)}%
\mathcal{P}_{jk]}^{(2)c}\right) \eta _{a}^{ijk}\eta ^{(2)b}  \nonumber \\
&&+2\left( \frac{\delta W_{ab}}{\delta \varphi _{c}}P_{c[i}^{(2)}\mathcal{P}%
_{jk]}^{(2)b}+W_{ab}\mathcal{P}_{ijk}^{b}\right) C^{aijk}.  \label{xx61}
\end{eqnarray}
The equation (\ref{xx49}) projected on antighost number one becomes 
\begin{equation}
\delta \stackrel{(2)}{\omega }_{1}+\gamma \stackrel{(1)}{\omega }%
_{1}=\partial ^{i}m_{i},  \label{xx62}
\end{equation}
which further yields 
\begin{eqnarray}
&&\stackrel{(1)}{\omega }_{1}=-\frac{\delta W_{ab}}{\delta \varphi _{c}}%
\left( P_{c}^{(2)j}C_{j}^{(2)a}\eta
^{(2)b}+P_{c[i}^{(2)}A_{j]}^{b}C^{aij}\right) -2W_{ab}C_{ij}^{a}\mathcal{P}%
^{(2)bij}  \nonumber \\
&&-\frac{1}{2}\frac{\delta M_{bc}^{a}}{\delta \varphi _{d}}\left(
P_{d}^{(2)j}B_{a0j}\eta ^{(2)b}\eta ^{(2)c}+P_{d[i}^{(2)}A_{j]}^{c}\eta
_{a}^{(2)ij}\eta ^{(2)b}+3P_{d}^{(2)i}\eta _{aijk}A^{bj}A^{ck}\right)  
\nonumber \\
&&-M_{bc}^{a}\left( \eta _{a}^{(2)ij}\eta ^{(2)b}\mathcal{P}%
_{ij}^{(2)c}-\eta _{a}^{ijk}A_{[i}^{b}\mathcal{P}_{jk]}^{(2)c}+\frac{1}{2}%
\mathcal{P}_{a}^{(2)}\eta ^{(2)b}\eta ^{(2)c}\right) .  \label{xx63}
\end{eqnarray}
Acting in the same manner in relation with the equation that governs the
antighost number zero element of $\omega _{1}$%
\begin{equation}
\delta \stackrel{(1)}{\omega }_{1}+\gamma \stackrel{(0)}{\omega }%
_{1}=\partial ^{i}l_{i},  \label{xx64}
\end{equation}
we arrive at the solution 
\begin{equation}
\stackrel{(0)}{\omega }_{1}=-W_{ab}\left( H_{0}^{a}\eta
^{(2)b}+C_{i}^{(2)a}A^{bi}\right) -M_{bc}^{a}\left( B_{a}^{0i}\eta
^{(2)b}A_{i}^{c}+\frac{1}{2}\eta _{a}^{(2)ij}A_{i}^{b}A_{j}^{c}\right) .
\label{xx65}
\end{equation}
In consequence, we succeeded in finding the complete form of the first-order
deformation of the BRST charge for the model under study.

\subsection{Higher-order deformations}

Next, we investigate the consistency of the first-order deformation,
described by the equation (\ref{xx43}). In view of this, by direct
computation we find that $[\Omega _{1},\ \Omega _{1}]=\int d^{3}x\Delta $,
with 
\begin{eqnarray}
&&\Delta =K^{abc}t_{abc}+ K_{d}^{abc}\frac{\delta t_{abc}}{\delta \varphi
_{d}}+K_{de}^{abc}\frac{\delta ^{2}t_{abc}}{\delta \varphi _{d}\delta
\varphi _{e}}+K_{def}^{abc}\frac{\delta ^{3}t_{abc}}{\delta \varphi
_{d}\delta \varphi _{e}\delta \varphi _{f}}  \nonumber \\
&&+U_{d}^{abc}t_{abc}^{d}+ U_{de}^{abc}\frac{\delta t_{abc}^{d}}{\delta
\varphi _{e}}+U_{def}^{abc}\frac{\delta ^{2}t_{abc}^{d}}{\delta \varphi
_{e}\delta \varphi _{f}}+U_{defg}^{abc}\frac{\delta ^{3}t_{abc}^{d}}{\delta
\varphi _{e}\delta \varphi _{f}\delta \varphi _{g}},  \label{xx66}
\end{eqnarray}
where we made the notations 
\begin{equation}
t_{abc}=W_{dc}M_{ab}^{d}+W_{da}\frac{\delta W_{bc}}{\delta \varphi _{d}}%
+W_{db}\frac{\delta W_{ca}}{\delta \varphi _{d}},  \label{xx67}
\end{equation}
\begin{equation}
t_{abc}^{d}=W_{e[a}\frac{\delta M_{bc]}^{d}}{\delta \varphi _{e}}%
+M_{e[a}^{d}M_{bc]}^{e},  \label{xx68}
\end{equation}
\begin{eqnarray}
&&K^{abc}=C^{aijk}\left( \eta ^{(2)b}\mathcal{P}_{ijk}^{c}-4\mathcal{P}%
_{[ij}^{(2)b}A_{k]}^{c}\right) +H_{0}^{a}\eta ^{(2)b}\eta
^{(2)c}-2C_{i}^{(2)a}\eta ^{(2)b}A^{ci}  \nonumber \\
&&+C_{ij}^{a}\left( 2A^{bi}A^{cj}+ \eta ^{(2)b}\mathcal{P}^{(2)cij}\right) ,
\label{xx69}
\end{eqnarray}
\begin{eqnarray}
&&K_{d}^{abc}=-2P_{d[i}^{(2)}A_{j]}^{c}C^{aij}\eta
^{(2)b}-6C_{ijk}^{a}P_{d}^{(2)i}A^{bj}A^{ck}+\frac{1}{3}C^{aijk}P_{d[i}^{(2)}%
\mathcal{P}_{jk]}^{(2)c}\eta ^{(2)b}  \nonumber \\
&&+\left(
P_{d}^{(2)i}C_{i}^{(2)a}-C_{ij}^{a}P_{d}^{ij}+P_{d}^{ijk}C_{ijk}^{a}\right)
\eta ^{(2)b}\eta ^{(2)c}+2C^{aijk}P_{d[ij}A_{k]}^{c}\eta ^{(2)b},
\label{xx70}
\end{eqnarray}
\begin{eqnarray}
&&K_{de}^{abc}=-\left(
C_{ij}^{a}P_{d}^{(2)i}P_{e}^{(2)j}+P_{d}^{(2)[i}P_{e}^{jk]}C_{ijk}^{a}%
\right) \eta ^{(2)b}\eta ^{(2)c}  \nonumber \\
&&+6P_{d}^{(2)i}P_{e}^{(2)j}C_{ijk}^{a}\eta ^{(2)b}A^{ck},  \label{xx71}
\end{eqnarray}
\begin{equation}
K_{def}^{abc}=-P_{d}^{(2)i}P_{e}^{(2)j}P_{f}^{(2)k}C_{ijk}^{a}\eta
^{(2)b}\eta ^{(2)c},  \label{xx72}
\end{equation}
\begin{eqnarray}
&&U_{d}^{abc}=\eta _{d}^{(2)ij}A_{i}^{a}A_{j}^{b}\eta ^{(2)c}-\eta
_{d}^{ijk}A_{i}^{a}A_{j}^{b}A_{k}^{c}+2\eta _{d}^{ijk}\eta ^{(2)b}\mathcal{P}%
_{[ij}^{(2)a}A_{k]}^{c}  \nonumber \\
&&+\left( \frac{1}{3}\mathcal{P}_{d}^{(2)}\eta
^{(2)c}-B_{d}^{0i}A_{i}^{c}+\eta _{d}^{(2)ij}\mathcal{P}_{ij}^{(2)c}+%
\mathcal{P}_{ijk}^{c}\eta _{d}^{ijk}\right) \eta ^{(2)a}\eta ^{(2)b},
\label{xx73}
\end{eqnarray}
\begin{eqnarray}
&&U_{de}^{abc}= \frac{1}{3!}\left( 2B_{d0i}P_{e}^{(2)i}-P_{e}^{ij}\eta
_{dij}^{(2)}+P_{e}^{ijk}\eta _{dijk}\right) \eta ^{(2)a}\eta ^{(2)b}\eta
^{(2)c}  \nonumber \\
&&+3P_{ei}^{(2)}\eta _{d}^{ijk}\eta ^{(2)c}A_{j}^{a}A_{k}^{b}  \nonumber \\
&&+\left( \frac{1}{2}\eta _{d}^{(2)ij}P_{e[i}^{(2)}A_{j]}^{c}+\eta
_{d}^{ijk}P_{e[i}^{(2)}\mathcal{P}_{jk]}^{(2)c}+\frac{1}{2}\eta
_{d}^{ijk}P_{e[ij}A_{k]}^{c}\right) \eta ^{(2)a}\eta ^{(2)b},  \label{xx74}
\end{eqnarray}
\begin{eqnarray}
&&U_{def}^{abc}=-\frac{1}{3!}\left( \eta
_{dij}^{(2)}P_{e}^{(2)i}P_{f}^{(2)j}+\eta
_{dijk}P_{e}^{(2)[i}P_{f}^{jk]}\right) \eta ^{(2)a}\eta ^{(2)b}\eta ^{(2)c} 
\nonumber \\
&&+\frac{3}{2}A^{ci}\eta _{dijk}P_{e}^{(2)j}P_{f}^{(2)k}\eta ^{(2)a}\eta
^{(2)b},  \label{xx75}
\end{eqnarray}
\begin{equation}
U_{defg}^{abc}=-\frac{1}{3!}\eta
_{dijk}P_{e}^{(2)i}P_{f}^{(2)j}P_{g}^{(2)k}\eta ^{(2)a}\eta ^{(2)b}\eta
^{(2)c}.  \label{xx76}
\end{equation}
The equation (\ref{xx43}) requires that $[\Omega _{1},\Omega _{1}]$ should
be $s$-exact. However, none of the terms in (\ref{xx66}) is so, hence $%
\Delta $ must vanish. This takes place if and only if 
\begin{equation}
t_{abc}=0,\;t_{abc}^{d}=0.  \label{sist}
\end{equation}
The solution to (\ref{sist}) reads as 
\begin{equation}
M_{ab}^{c}=\frac{\delta W_{ab}}{\delta \varphi _{c}},  \label{sol}
\end{equation}
where, in addition, the antisymmetric functions $W_{ab}$ fulfill the
identity 
\begin{equation}
W_{e[a}\frac{\delta W_{bc]}}{\delta \varphi _{e}}=0.  \label{xx77}
\end{equation}

In consequence, the consistency of the first-order deformation of the BRST
charge implies that the functions $W_{ab}$ are antisymmetric and check
Jacobi's identity (\ref{xx77}) corresponding to an open (nonlinear) algebra.
Further, we can take $\Omega _{2}=0$, the remaining higher-order deformation
equations being satisfied with the choice 
\begin{equation}
\Omega _{k}=0,\;k>2.  \label{xx78}
\end{equation}
This completes the approach to the deformed BRST charge of the free model
under discussion.

\section{Deformed BRST-invariant Hamiltonian}

In order to analyse the deformation of the BRST-invariant Hamiltonian (\ref
{xx27}), we initially solve the equation (\ref{xx47}). By direct computation
we get that 
\begin{eqnarray}
&&\left[ \Omega _{1},H_{0\mathrm{B}}\right] =\int d^{3}x\left( -\left( \frac{%
\delta W_{ab}}{\delta \varphi _{c}}P_{c}^{ijk}-\frac{\delta ^{2}W_{ab}}{%
\delta \varphi _{c}\delta \varphi _{d}}P_{c}^{(2)[i}P_{d}^{jk]}\right.
\right.  \nonumber \\
&&\left. -\frac{\delta ^{3}W_{ab}}{\delta \varphi _{c}\delta \varphi
_{d}\delta \varphi _{e}}P_{c}^{(2)i}P_{d}^{(2)j}P_{e}^{(2)k}\right)
C_{ijk}^{a}\eta ^{(1)b}-\frac{1}{2}\left( \frac{\delta M_{bc}^{a}}{\delta
\varphi _{d}}P_{d}^{ijk}\right.  \nonumber \\
&&\left. -\frac{\delta ^{2}M_{bc}^{a}}{\delta \varphi _{d}\delta \varphi _{e}%
}P_{d}^{(2)[i}P_{e}^{jk]}-\frac{\delta ^{3}M_{bc}^{a}}{\delta \varphi
_{d}\delta \varphi _{e}\delta \varphi _{f}}%
P_{d}^{(2)i}P_{e}^{(2)j}P_{f}^{(2)k}\right) \eta _{aijk}\eta ^{(2)b}\eta
^{(1)c}  \nonumber \\
&&+\left( \frac{\delta W_{ab}}{\delta \varphi _{c}}P_{c}^{ij}+\frac{\delta
^{2}W_{ab}}{\delta \varphi _{c}\delta \varphi _{d}}P_{c}^{(2)i}P_{d}^{(2)j}%
\right) \left( C_{ij}^{a}\eta ^{(1)b}+3C_{ijk}^{a}\partial
^{k}A_{0}^{b}\right)  \nonumber \\
&&+\frac{1}{2}\left( \frac{\delta M_{bc}^{a}}{\delta \varphi _{d}}P_{d}^{ij}+%
\frac{\delta ^{2}M_{bc}^{a}}{\delta \varphi _{d}\delta \varphi _{e}}%
P_{d}^{(2)i}P_{e}^{(2)j}\right) \left( \eta _{aij}^{(2)}\eta ^{(2)b}\eta
^{(1)c}+\frac{1}{2}\eta _{aij}^{(1)}\eta ^{(2)b}\eta ^{(2)c}\right. 
\nonumber \\
&&\left. +3\eta _{aijk}\eta ^{(2)b}\partial ^{k}A_{0}^{c}+3\eta _{aijk}\eta
^{(1)b}A^{ck}\right)  \nonumber \\
&&+3\frac{\delta M_{bc}^{a}}{\delta \varphi _{d}}P_{d}^{(2)i}\mathcal{P}%
^{(2)cjk}\eta _{aijk}\eta ^{(1)b}+M_{bc}^{a}\left( \eta _{a}^{ijk}\eta
^{(1)b}\mathcal{P}_{ijk}^{c}-\mathcal{P}^{(2)a}\eta ^{(2)b}\eta
^{(1)c}\right)  \nonumber \\
&&-\frac{\delta W_{ab}}{\delta \varphi _{c}}P_{c}^{(2)i}\left(
C_{i}^{(1)a}\eta ^{(2)b}+C_{i}^{(2)a}\eta ^{(1)b}+2\partial
^{j}A_{0}^{b}C_{ij}^{a}\right)  \nonumber \\
&&-\frac{\delta M_{bc}^{a}}{\delta \varphi _{d}}P_{d}^{(2)i}\left(
B_{a0i}\eta ^{(2)b}\eta ^{(1)c}+\frac{1}{2}\partial ^{j}B_{aij}\eta
^{(2)b}\eta ^{(2)c}+A^{cj}\eta _{aij}^{(2)}\eta ^{(1)b}\right.  \nonumber \\
&&\left. +A^{cj}\eta _{aij}^{(1)}\eta ^{(2)b}+\partial ^{j}A_{0}^{c}\eta
_{aij}^{(2)}\eta ^{(2)b}+3\eta _{aijk}A^{bj}\partial ^{k}A_{0}^{c}\right) 
\nonumber \\
&&-M_{bc}^{a}\left( \eta _{a}^{(1)ij}\eta ^{(2)b}\mathcal{P}%
_{ij}^{(2)c}+\eta _{a}^{(2)ij}\eta ^{(1)b}\mathcal{P}_{ij}^{(2)c}-3\eta
_{a}^{ijk}\partial _{i}A_{0}^{b}\mathcal{P}_{jk}^{(2)c}\right)  \nonumber \\
&&-W_{ab}\left( H_{0}^{a}\eta ^{(1)b}-\partial ^{i}H_{i}^{a}\eta
^{(2)b}+C_{i}^{(1)a}A^{bi}+C_{i}^{(2)a}\partial ^{i}A_{0}^{b}\right) 
\nonumber \\
&&-M_{bc}^{a}\left( B_{a}^{0i}\eta ^{(2)b}\partial
_{i}A_{0}^{c}+B_{a}^{0i}\eta ^{(1)b}A_{i}^{c}-\partial _{i}B_{a}^{ij}\eta
^{(2)b}A_{j}^{c}\right.  \nonumber \\
&&\left. \left. +\eta _{a}^{(2)ij}A_{i}^{b}\partial _{j}A_{0}^{c}+\frac{1}{2}%
\eta _{a}^{(1)ij}A_{i}^{b}A_{j}^{c}\right) \right) ,  \label{xx79}
\end{eqnarray}
which offers us the first-order deformation of the BRST-invariant
Hamiltonian as 
\begin{eqnarray}
&&h_{1}=-\left( \frac{\delta W_{ab}}{\delta \varphi _{c}}P_{c}^{ijk}-\frac{%
\delta ^{2}W_{ab}}{\delta \varphi _{c}\delta \varphi _{d}}%
P_{c}^{(2)[i}P_{d}^{jk]}\right.  \nonumber \\
&&\left. -\frac{\delta ^{3}W_{ab}}{\delta \varphi _{c}\delta \varphi
_{d}\delta \varphi _{e}}P_{c}^{(2)i}P_{d}^{(2)j}P_{e}^{(2)k}\right)
C_{ijk}^{a}A_{0}^{b}  \nonumber \\
&&-\frac{1}{2}\left( \frac{\delta M_{bc}^{a}}{\delta \varphi _{d}}%
P_{d}^{ijk}-\frac{\delta ^{2}M_{bc}^{a}}{\delta \varphi _{d}\delta \varphi
_{e}}P_{d}^{(2)[i}P_{e}^{jk]}\right.  \nonumber \\
&&\left. -\frac{\delta ^{3}M_{bc}^{a}}{\delta \varphi _{d}\delta \varphi
_{e}\delta \varphi _{f}}P_{d}^{(2)i}P_{e}^{(2)j}P_{f}^{(2)k}\right) \eta
_{aijk}\eta ^{(2)b}A_{0}^{c}  \nonumber \\
&&+\left( \frac{\delta W_{ab}}{\delta \varphi _{c}}P_{c}^{ij}+\frac{\delta
^{2}W_{ab}}{\delta \varphi _{c}\delta \varphi _{d}}P_{c}^{(2)i}P_{d}^{(2)j}%
\right) C_{ij}^{a}A_{0}^{b}  \nonumber \\
&&+\frac{\delta W_{ab}}{\delta \varphi _{c}}P_{c}^{(2)i}\left( H_{i}^{a}\eta
^{(2)b}-C_{i}^{(2)a}A_{0}^{b}\right)  \nonumber \\
&&+\frac{1}{2}\left( \frac{\delta M_{bc}^{a}}{\delta \varphi _{d}}P_{d}^{ij}+%
\frac{\delta ^{2}M_{bc}^{a}}{\delta \varphi _{d}\delta \varphi _{e}}%
P_{d}^{(2)i}P_{e}^{(2)j}\right) \times  \nonumber \\
&&\left( \eta _{aij}^{(2)}\eta ^{(2)b}A_{0}^{c}+\frac{1}{2}B_{aij}\eta
^{(2)b}\eta ^{(2)c}+3\eta _{aijk}A_{0}^{b}A^{ck}\right)  \nonumber \\
&&+\frac{\delta M_{bc}^{a}}{\delta \varphi _{d}}\left( P_{d[i}^{(2)}\mathcal{%
P}_{jk]}^{(2)c}\eta _{a}^{ijk}A_{0}^{b}-P_{d}^{(2)i}B_{a0i}\eta
^{(2)b}A_{0}^{c}\right.  \nonumber \\
&&\left. +\frac{1}{2}P_{d[i}^{(2)}A_{j]}^{c}B_{a}^{ij}\eta ^{(2)b}-\frac{1}{2%
}P_{d[i}^{(2)}A_{j]}^{c}\eta _{a}^{(2)ij}A_{0}^{b}\right)  \nonumber \\
&&+M_{bc}^{a}\left( \eta _{a}^{(2)ij}A_{0}^{b}\mathcal{P}_{ij}^{(2)c}+\eta
_{a}^{ijk}A_{0}^{b}\mathcal{P}_{ijk}^{c}\right.  \nonumber \\
&&\left. -\mathcal{P}_{a}^{(2)}\eta ^{(2)b}A_{0}^{c}-B_{a}^{ij}\eta ^{(2)b}%
\mathcal{P}_{ij}^{(2)c}\right)  \nonumber \\
&&-W_{ab}H_{\mu }^{a}A^{b\mu }-\frac{1}{2}M_{bc}^{a}B_{a}^{\mu \nu }A_{\mu
}^{b}A_{\nu }^{c}.  \label{xx80}
\end{eqnarray}
In (\ref{xx79}--\ref{xx80}) and further the functions $M_{bc}^{a}$ are
expressed as in (\ref{sol}).

Now, we pass to the equation (\ref{xx48}). After some computation, we infer
that $[\Omega _{1},\ H_{1}]=\int d^{3}x\overline{\Delta }$, where 
\begin{eqnarray}
&&\overline{\Delta }= \overline{K}^{abc}t_{abc}+\overline{K}_{d}^{abc}\frac{%
\delta t_{abc}}{\delta \varphi _{d}}+\overline{K}_{de}^{abc}\frac{\delta
^{2}t_{abc}}{\delta \varphi _{d}\delta \varphi _{e}}+\overline{K}_{def}^{abc}%
\frac{\delta ^{3}t_{abc}}{\delta \varphi _{d}\delta \varphi _{e}\delta
\varphi _{f}}  \nonumber \\
&&+\overline{U}_{d}^{abc}t_{abc}^{d}+\overline{U}_{de}^{abc}\frac{\delta
t_{abc}^{d}}{\delta \varphi _{e}}+\overline{U}_{def}^{abc}\frac{\delta
^{2}t_{abc}^{d}}{\delta \varphi _{e}\delta \varphi _{f}}+\overline{U}%
_{defg}^{abc}\frac{\delta ^{3}t_{abc}^{d}}{\delta \varphi _{e}\delta \varphi
_{f}\delta \varphi _{g}},  \label{xx81}
\end{eqnarray}
where the following notations were employed 
\begin{equation}
\overline{K}^{abc}=H_{\mu }^{a}\eta ^{(2)b}A^{c\mu
}-C_{i}^{(2)a}A_{0}^{b}A^{ci}-2\left( C_{ij}^{a}A_{0}^{b}\mathcal{P}%
^{(2)cij}-C_{ijk}^{a}A_{0}^{b}\mathcal{P}^{cijk}\right) ,  \label{xx82}
\end{equation}
\begin{eqnarray}
&&\overline{K}_{d}^{abc}=\left(
P_{d}^{(2)i}C_{i}^{(2)a}-P_{d}^{ij}C_{ij}^{a}+P_{d}^{ijk}C_{ijk}^{a}\right)
\eta ^{(2)b}A_{0}^{c}+\frac{1}{2}P_{d}^{(2)i}H_{i}^{a}\eta ^{(2)b}\eta
^{(2)c}  \nonumber \\
&&+\left(
C^{aijk}P_{d[ij}A_{k]}^{c}-C^{aij}P_{d[i}^{(2)}A_{j]}^{c}+2P_{d[i}^{(2)}%
\mathcal{P}_{jk]}^{(2)c}C^{aijk}\right) A_{0}^{b},  \label{xx83}
\end{eqnarray}
\begin{equation}
\overline{K}_{de}^{abc}=P_{d}^{(2)i}P_{e}^{(2)j}\left(
3C_{ijk}^{a}A_{0}^{b}A^{ck}-C_{ij}^{a}\eta ^{(2)b}A_{0}^{c}\right)
-P_{d}^{(2)[i}P_{e}^{jk]}C_{ijk}^{a}\eta ^{(2)b}A_{0}^{c},  \label{xx84}
\end{equation}
\begin{equation}
\overline{K}_{def}^{abc}=-P_{d}^{(2)i}P_{e}^{(2)j}P_{f}^{(2)k}C_{ijk}^{a}%
\eta ^{(2)b}A_{0}^{c},  \label{xx85}
\end{equation}
\begin{eqnarray}
&&\overline{U}_{d}^{abc}= \frac{1}{2}\left( \eta
_{d}^{(2)ij}A_{i}^{a}A_{j}^{b}+\mathcal{P}_{d}^{(2)}\eta ^{(2)a}\eta
^{(2)b}+2\eta _{d}^{ijk}\left( \mathcal{P}_{[ij}^{(2)a}A_{k]}^{b}+\eta
^{(2)b}\mathcal{P}_{ijk}^{a}\right) \right) A_{0}^{c}  \nonumber \\
&&-\left( \frac{1}{2}B_{d}^{ij} \eta ^{(2)a}\eta ^{(2)b}+\eta
_{d}^{(2)ij}\eta ^{(2)a}A_{0}^{b}\right) \mathcal{P}_{ij}^{(2)c}-\frac{1}{2}%
B_{d}^{\mu \nu }A_{\mu }^{a}A_{\nu }^{b}\eta ^{(2)c},  \label{xx86}
\end{eqnarray}
\begin{eqnarray}
&&\overline{U}_{de}^{abc}=\frac{1}{4}\left(
2P_{e}^{(2)i}B_{d0i}-P_{e}^{ij}\eta _{dij}^{(2)}+P_{e}^{ijk}\eta
_{dijk}\right) \eta ^{(2)a}\eta ^{(2)b}A_{0}^{c}  \nonumber \\
&&+\left( \frac{1}{3!}P_{eij}\eta ^{(2)c}- \frac{1}{4}%
P_{e[i}^{(2)}A_{j]}^{c}\right) B_{d}^{ij}\eta ^{(2)a}\eta ^{(2)b}+\eta
_{d}^{ijk}P_{e[i}^{(2)}\mathcal{P}_{jk]}^{(2)c}\eta ^{(2)a}A_{0}^{b} 
\nonumber \\
&&+\frac{1}{2}\left( P_{e[i}^{(2)}A_{j]}^{b}\eta
_{d}^{(2)ij}-P_{e[ij}A_{k]}^{b}\eta _{d}^{ijk}\right) \eta ^{(2)a}A_{0}^{c}+%
\frac{3}{2}P_{ei}^{(2)}\eta _{d}^{ijk}A_{j}^{a}A_{k}^{b}A_{0}^{c},
\label{xx87}
\end{eqnarray}
\begin{eqnarray}
&&\overline{U}_{def}^{abc}= -\frac{1}{4}\left( P_{e}^{(2)i}P_{f}^{(2)j}\eta
_{dij}^{(2)}+P_{e}^{(2)[i}P_{f}^{jk]}\eta _{dijk}\right) \eta ^{(2)a}\eta
^{(2)b}A_{0}^{c}  \nonumber \\
&&+P_{e}^{(2)i}P_{f}^{(2)j}\left( \frac{1}{3!}B_{dij}\eta ^{(2)a}\eta
^{(2)b}\eta ^{(2)c}-\frac{3}{2}\eta _{dijk}\eta
^{(2)a}A^{bk}A_{0}^{c}\right) ,  \label{xx88}
\end{eqnarray}
\begin{equation}
\overline{U}_{defg}^{abc}=-\frac{1}{4}P_{e}^{(2)i}P_{f}^{(2)j}P_{g}^{(2)k}%
\eta _{dijk}\eta ^{(2)a}\eta ^{(2)b}A_{0}^{c}.  \label{xx89}
\end{equation}
Using (\ref{sist}), we find that $\overline{\Delta }=0$, hence we can set $%
H_{2}=0$. The remaining equations are then satisfied for 
\begin{equation}
H_{k}=0,\;k>2.  \label{xx90}
\end{equation}
In this way, the deformed BRST-invariant Hamiltonian is also completely
computed.

\section{Interacting theory}

At this point, we are in the position to identify the deformed gauge theory
corresponding to the starting free model. Putting together the results
deduced so far, it follows that the complete expression of the deformed BRST
charge consistent to all orders in the deformation parameter is 
\begin{eqnarray}
&&\Omega =\int d^{3}x\left( \eta ^{(1)a}\pi
_{a}^{0}+C_{i}^{(1)a}p_{a}^{i}+\eta _{a}^{(1)ij}\pi
_{ij}^{a}-C_{i}^{(2)a}\left( \partial ^{i}\varphi _{a}+gW_{ab}A^{bi}\right)
\right.  \nonumber \\
&&+\eta ^{(2)a}\left( - \partial _{i}B_{a}^{0i}-g\frac{\delta W_{ab}}{\delta
\varphi _{c}}B_{c}^{0i}A_{i}^{b}+gW_{ab}H_{0}^{b}\right)  \nonumber \\
&&-\frac{1}{2}\eta _{a}^{(2)ij}\left( \partial _{[i}A_{j]}^{a}+ g\frac{%
\delta W_{bc}}{\delta \varphi _{a}}A_{i}^{b}A_{j}^{c}\right)  \nonumber \\
&&+C_{ij}^{a}\left( \partial ^{[i}P_{a}^{(2)j]}- g\frac{\delta W_{ab}}{%
\delta \varphi _{c}}P_{c}^{(2)[i}A^{bj]}-2gW_{ab}\mathcal{P}^{(2)bij}\right)
\nonumber \\
&&+\eta _{a}^{ijk}\left( \partial _{[i} \mathcal{P}_{jk]}^{(2)a}-g\frac{%
\delta W_{bc}}{\delta \varphi _{a}}A_{[i}^{c}\mathcal{P}_{jk]}^{(2)b}-\frac{3%
}{2}g\frac{\delta ^{2}W_{bc}}{\delta \varphi _{a}\delta \varphi _{d}}%
P_{di}^{(2)}A_{j}^{b}A_{k}^{c}\right)  \nonumber \\
&&-g\frac{\delta W_{bc}}{\delta \varphi _{a}}\left( \eta _{a}^{(2)ij}\eta
^{(2)b}\mathcal{P}_{ij}^{(2)c}+\frac{1}{2}\mathcal{P}_{a}^{(2)}\eta
^{(2)b}\eta ^{(2)c}+P_{a}^{(2)i}C_{i}^{(2)b}\eta ^{(2)c}\right)  \nonumber \\
&&-\frac{g}{2}\frac{\delta ^{2} W_{bc}}{\delta \varphi _{a}\delta \varphi
_{d}}\left( P_{d}^{(2)i}B_{a0i}\eta ^{(2)b}\eta
^{(2)c}+P_{d[i}^{(2)}A_{j]}^{c}\eta _{a}^{(2)ij}\eta ^{(2)b}\right) 
\nonumber \\
&&+C_{ijk}^{a}\left( \partial ^{[i}P_{a}^{jk]}+ g\frac{\delta W_{ab}}{\delta
\varphi _{c}}A^{b[i}P_{c}^{jk]}+2gW_{ab}\mathcal{P}^{bijk}\right.  \nonumber
\\
&&\left. +3g\frac{\delta ^{2}W_{ab}}{\delta \varphi _{c}\delta \varphi _{d}}%
P_{c}^{(2)i}P_{d}^{(2)j}A^{bk}+2g\frac{\delta W_{ab}}{\delta \varphi _{c}}%
P_{c}^{(2)[i}\mathcal{P}^{(2)bjk]}\right)  \nonumber \\
&&+g\left( \frac{\delta W_{bc}}{\delta \varphi _{a}}\mathcal{P}_{ijk}^{c}+%
\frac{\delta ^{2}W_{bc}}{\delta \varphi _{a}\delta \varphi _{d}}P_{d[i}^{(2)}%
\mathcal{P}_{jk]}^{(2)c}\right) \eta _{a}^{ijk}\eta ^{(2)b}  \nonumber \\
&&+g\left( \frac{\delta W_{bc}}{\delta \varphi _{a}}P_{a}^{ij}+ \frac{\delta
^{2}W_{bc}}{\delta \varphi _{a}\delta \varphi _{d}}P_{a}^{(2)i}P_{d}^{(2)j}%
\right) C_{ij}^{b}\eta ^{(2)c}  \nonumber \\
&&+\frac{g}{2}\left( \frac{\delta ^{2}W_{bc}}{\delta \varphi _{a}\delta
\varphi _{d}}P_{d}^{ij}+\frac{\delta ^{3}W_{bc}}{\delta \varphi _{a}\delta
\varphi _{d}\delta \varphi _{f}}P_{d}^{(2)i}P_{f}^{(2)j}\right) \left( \frac{%
1}{2}\eta _{aij}^{(2)}\eta ^{(2)b}\eta ^{(2)c}\right.  \nonumber \\
&&\left. +3\eta _{aijk}\eta ^{(2)b}A^{ck}\right) -\frac{g}{4}\left( \frac{%
\delta ^{2}W_{bc}}{\delta \varphi _{d}\delta \varphi _{a}}P_{d}^{ijk}-\frac{%
\delta ^{3}W_{bc}}{\delta \varphi _{d}\delta \varphi _{e}\delta \varphi _{a}}%
P_{d}^{(2)[i}P_{e}^{jk]}\right.  \nonumber \\
&&\left. -\frac{\delta ^{4}W_{bc}}{\delta \varphi _{d}\delta \varphi
_{e}\delta \varphi _{f}\delta \varphi _{a}}%
P_{d}^{(2)i}P_{e}^{(2)j}P_{f}^{(2)k}\right) \eta _{aijk}\eta ^{(2)b}\eta
^{(2)c}  \nonumber \\
&&-g\left( \frac{\delta W_{ab}}{\delta \varphi _{c}}P_{c}^{ijk}-\frac{\delta
^{2}W_{ab}}{\delta \varphi _{c}\delta \varphi _{d}}P_{c}^{(2)[i}P_{d}^{jk]}%
\right.  \nonumber \\
&&\left. \left. -\frac{\delta ^{3}W_{ab}}{\delta \varphi _{c}\delta \varphi
_{d}\delta \varphi _{e}}P_{c}^{(2)i}P_{d}^{(2)j}P_{e}^{(2)k}\right)
C_{ijk}^{a}\eta ^{(2)b}\right) ,  \label{xx92}
\end{eqnarray}
while that of the BRST-invariant Hamiltonian can be written in the form 
\begin{eqnarray}
&&H_{\mathrm{B}}= \int d^{3}x\left( -H_{i}^{a}\left( \partial ^{i}\varphi
_{a}+gW_{ab}A^{bi}\right) \right.  \nonumber \\
&&-\frac{1}{2}B_{a}^{ij}\left( \partial _{[i}A_{j]}^{a}+ g\frac{\delta W_{bc}%
}{\delta \varphi _{a}}A_{i}^{b}A_{j}^{c}\right)  \nonumber \\
&&+A_{0}^{a}\left( -\partial _{i}B_{a}^{0i}- g\frac{\delta W_{ab}}{\delta
\varphi _{c}}B_{c}^{0i}A_{i}^{b}+gW_{ab}H_{0}^{b}\right)  \nonumber \\
&&+\eta ^{(1)a}\mathcal{P}_{a}^{(2)}+C_{i}^{(1)a}P_{a}^{(2)i}+ \eta
_{a}^{(1)ij}\mathcal{P}_{ij}^{(2)a}  \nonumber \\
&&-g\frac{\delta W_{bc}}{\delta \varphi _{a}}\left( \mathcal{P}%
_{a}^{(2)}\eta ^{(2)b}A_{0}^{c}+B_{a}^{ij}\eta ^{(2)b}\mathcal{P}%
_{ij}^{(2)c}-\eta _{a}^{(2)ij}A_{0}^{b}\mathcal{P}_{ij}^{(2)c}\right) 
\nonumber \\
&&-g\frac{\delta ^{2}W_{bc}}{\delta \varphi _{a}\delta \varphi _{d}}%
P_{d}^{(2)i}\left( B_{a0i}\eta ^{(2)b}A_{0}^{c}+\eta
_{a}^{(2)ij}A_{j}^{c}A_{0}^{b}-B_{a}^{ij}A_{j}^{c}\eta ^{(2)b}\right) 
\nonumber \\
&&+g\frac{\delta W_{ab}}{\delta \varphi _{c}}P_{c}^{(2)i} \left(
H_{i}^{a}\eta ^{(2)b}-C_{i}^{(2)a}A_{0}^{b}\right)  \nonumber \\
&&+g\left( \frac{\delta W_{bc}}{\delta \varphi _{a}}\mathcal{P}_{ijk}^{c}+%
\frac{\delta ^{2}W_{bc}}{\delta \varphi _{a}\delta \varphi _{d}}P_{d[i}^{(2)}%
\mathcal{P}_{jk]}^{(2)c}\right) \eta _{a}^{ijk}A_{0}^{b}  \nonumber \\
&&+\frac{g}{2}\left( \frac{\delta ^{2}W_{bc}}{\delta \varphi _{a}\delta
\varphi _{d}}P_{d}^{ij}+\frac{\delta ^{3}W_{bc}}{\delta \varphi _{a}\delta
\varphi _{d}\delta \varphi _{f}}P_{d}^{(2)i}P_{f}^{(2)j}\right) \times 
\nonumber \\
&&\left( \eta _{aij}^{(2)}\eta ^{(2)b}A_{0}^{c}+\frac{1}{2}B_{aij}\eta
^{(2)b}\eta ^{(2)c}+3\eta _{aijk}A_{0}^{b}A^{ck}\right)  \nonumber \\
&&+g\left( \frac{\delta W_{ab}}{\delta \varphi _{c}}P_{c}^{ij}+\frac{\delta
^{2}W_{ab}}{\delta \varphi _{c}\delta \varphi _{d}}P_{c}^{(2)i}P_{d}^{(2)j}%
\right) C_{ij}^{a}A_{0}^{b}  \nonumber \\
&&-\frac{g}{2}\left( \frac{\delta ^{2}W_{bc}}{\delta \varphi _{d}\delta
\varphi _{a}}P_{d}^{ijk}-\frac{\delta ^{3}W_{bc}}{\delta \varphi _{d}\delta
\varphi _{e}\delta \varphi _{a}}P_{d}^{(2)[i}P_{e}^{jk]}\right.  \nonumber \\
&&\left. -\frac{\delta ^{4}W_{bc}}{\delta \varphi _{d}\delta \varphi
_{e}\delta \varphi _{f}\delta \varphi _{a}}%
P_{d}^{(2)i}P_{e}^{(2)j}P_{f}^{(2)k}\right) \eta _{aijk}\eta ^{(2)b}A_{0}^{c}
\nonumber \\
&&-g\left( \frac{\delta W_{ab}}{\delta \varphi _{c}}P_{c}^{ijk}-\frac{\delta
^{2}W_{ab}}{\delta \varphi _{c}\delta \varphi _{d}}P_{c}^{(2)[i}P_{d}^{jk]}%
\right.  \nonumber \\
&&\left. \left. -\frac{\delta ^{3}W_{ab}}{\delta \varphi _{c}\delta \varphi
_{d}\delta \varphi _{e}}P_{c}^{(2)i}P_{d}^{(2)j}P_{e}^{(2)k}\right)
C_{ijk}^{a}A_{0}^{b}\right) .  \label{xx93}
\end{eqnarray}

From the terms of antighost number zero in (\ref{xx92}) we see that only the
secondary constraints are deformed as 
\begin{equation}
\overline{G}_{a}^{(2)}\equiv -\left( D_{i}\right)
_{a}^{\;\;b}B_{b}^{0i}+gW_{ab}H_{0}^{b}\approx 0,  \label{xx94}
\end{equation}
\begin{equation}
\overline{G}_{ij}^{(2)a}\equiv -\frac{1}{2}\overline{F}_{ij}^{a}\approx 0,
\label{xx95}
\end{equation}
\begin{equation}
\overline{\gamma }_{a}^{(2)i}\equiv -D^{i}\varphi _{a}\approx 0,
\label{xx96}
\end{equation}
where we employed the notations 
\begin{equation}
\left( D_{i}\right) _{a}^{\;\;b}=\delta _{a}^{\;\;b}\partial _{i}+g\frac{%
\delta W_{ac}}{\delta \varphi _{b}}A_{i}^{c}  \label{xx97}
\end{equation}
\begin{equation}
\overline{F}_{ij}^{a}=\partial _{[i}A_{j]}^{a}+g\frac{\delta W_{bc}}{\delta
\varphi _{a}}A_{i}^{b}A_{j}^{c},  \label{xx98}
\end{equation}
\begin{equation}
D^{i}\varphi _{a}=\partial ^{i}\varphi _{a}+gW_{ab}A^{bi}.  \label{xx99}
\end{equation}
The pieces linear in the antighost number one antighosts show that some of
the Dirac brackets among the new constraint functions are modified as 
\begin{equation}
\left[ \overline{G}_{a}^{(2)},\overline{G}_{b}^{(2)}\right] =-g\left( \frac{%
\delta W_{ab}}{\delta \varphi _{c}}\overline{G}_{c}^{(2)}+\frac{\delta
^{2}W_{ab}}{\delta \varphi _{c}\delta \varphi _{d}}\overline{\gamma }%
_{c}^{(2)i}B_{d0i}\right) ,  \label{xx100}
\end{equation}
\begin{equation}
\left[ \overline{G}_{a}^{(2)},\overline{G}_{ij}^{(2)b}\right] =g\left( \frac{%
\delta W_{ac}}{\delta \varphi _{b}}\overline{G}_{ij}^{(2)c}+\frac{1}{2}\frac{%
\delta ^{2}W_{ac}}{\delta \varphi _{b}\delta \varphi _{d}}\overline{\gamma }%
_{d[i}^{(2)}A_{j]}^{c}\right) ,  \label{xx101}
\end{equation}
\begin{equation}
\left[ \overline{G}_{a}^{(2)},\overline{\gamma }_{b}^{(2)i}\right] =-g\frac{%
\delta W_{ab}}{\delta \varphi _{c}}\overline{\gamma }_{c}^{(2)i}.
\label{xx102}
\end{equation}
The elements linear in the antighost number two antighosts yield the
deformed first-stage reducibility relations under the form 
\begin{equation}
\overline{Z}_{bklm}^{aij}\overline{G}_{ij}^{(2)b}+\overline{Z}_{klmi}^{ab}%
\overline{\gamma }_{b}^{(2)i}=0,  \label{xx103}
\end{equation}
\begin{equation}
\overline{Z}_{ab}^{ijkl}\overline{G}_{kl}^{(2)b}+\overline{Z}_{ak}^{bij}%
\overline{\gamma }_{b}^{(2)k}=0,  \label{xx104}
\end{equation}
where the associated first-stage reducibility functions read as 
\begin{equation}
\overline{Z}_{bklm}^{aij}=\frac{1}{2}\left( D_{[k}\right) _{\;\;b}^{a}\delta
_{l}^{i}\delta _{m]}^{j},\;\overline{Z}_{klmi}^{ab}=-\frac{g}{4}\frac{\delta
^{2}W_{cd}}{\delta \varphi _{a}\delta \varphi _{b}}%
g_{i[k}A_{l}^{c}A_{m]}^{d},  \label{xx105}
\end{equation}
\begin{equation}
\overline{Z}_{ab}^{ijkl}=-gW_{ab}\left( g^{ik}g^{jl}-g^{il}g^{jk}\right) ,\;%
\overline{Z}_{ak}^{bij}=\left( D^{[i}\right) _{a}^{\;\;b}\delta _{k}^{j]},
\label{xx106}
\end{equation}
with 
\begin{equation}
\left( D_{i}\right) _{\;\;b}^{a}=\delta _{\;\;b}^{a}\partial _{i}-g\frac{%
\delta W_{bc}}{\delta \varphi _{a}}A_{i}^{c}.  \label{xx107}
\end{equation}
The antighost number three terms underline that the second-stage
reducibility relations 
\begin{eqnarray}
&&\overline{Z}_{aj_{1}j_{2}}^{bi_{1}i_{2}i_{3}}\overline{Z}%
_{bk}^{cj_{1}j_{2}}f_{c}^{k}+\overline{Z}%
_{ab}^{i_{1}i_{2}i_{3}j_{1}j_{2}j_{3}}\overline{Z}%
_{j_{1}j_{2}j_{3}k}^{bc}f_{c}^{k}=  \nonumber \\
&&-g\left( 2\frac{\delta W_{ab}}{\delta \varphi _{c}}\overline{G}%
^{(2)b[i_{1}i_{2}}f_{c}^{i_{3}]}+\frac{\delta ^{2}W_{ab}}{\delta \varphi
_{c}\delta \varphi _{d}}\overline{\gamma }%
_{c}^{(2)[i_{1}}A^{bi_{2}}f_{d}^{i_{3}]}\right) ,  \label{xx108}
\end{eqnarray}
\begin{eqnarray}
&&\overline{Z}_{aj_{1}j_{2}}^{bi_{1}i_{2}i_{3}}\overline{Z}%
_{bc}^{j_{1}j_{2}kl}f_{kl}^{c}+\overline{Z}%
_{ab}^{i_{1}i_{2}i_{3}j_{1}j_{2}j_{3}}\overline{Z}%
_{cj_{1}j_{2}j_{3}}^{bkl}f_{kl}^{c}=  \nonumber \\
&&2g\frac{\delta W_{ab}}{\delta \varphi _{c}}f^{b[i_{1}i_{2}}\overline{%
\gamma }_{c}^{(2)i_{3}]},  \label{xx109}
\end{eqnarray}
hold on-shell, where $f_{c}^{k}$ and $f_{kl}^{c}$ are arbitrary smooth
functions, the latter being antisymmetric in their spatial indices, where
the second-stage reducibility functions are given by 
\begin{equation}
\overline{Z}_{aj_{1}j_{2}}^{bi_{1}i_{2}i_{3}}=\frac{1}{2}\left(
D^{[i_{1}}\right) _{a}^{\;\;b}\delta _{j_{1}}^{i_{2}}\delta
_{j_{2}}^{i_{3}]},  \label{xx110}
\end{equation}
\begin{equation}
\overline{Z}_{ab}^{i_{1}i_{2}i_{3}j_{1}j_{2}j_{3}}=\frac{g}{3}W_{ab}\left(
\sum\limits_{\sigma \in S_{3}}(-)^{\sigma }g^{i_{1}j_{\sigma
(1)}}g^{i_{2}j_{\sigma (2)}}g^{i_{3}j_{\sigma (3)}}\right) .  \label{xx111}
\end{equation}
In (\ref{xx111}), $S_{3}$ signifies the set of permutations of $\left\{
1,2,3\right\} $, and $(-)^{\sigma }$ means the parity of the permutation $%
\sigma $ pertaining to $S_{3}$. Now, we investigate the modified
BRST-invariant Hamiltonian (\ref{xx93}). The component of antighost number
zero 
\begin{equation}
H=\int d^{3}x\left( H_{i}^{a}\overline{\gamma }_{a}^{(2)i}+B_{a}^{ij}%
\overline{G}_{ij}^{(2)a}+A_{0}^{a}\overline{G}_{a}^{(2)}\right) ,
\label{xx112}
\end{equation}
represents nothing but the new first-class Hamiltonian, while the terms
linear in the antighost number one antighosts give the deformed gauge
algebra relations 
\begin{equation}
\left[ H,G_{a}^{(1)}\right] =\overline{G}_{a}^{(2)},  \label{xx113}
\end{equation}
\begin{eqnarray}
&&\left[ H,\overline{G}_{a}^{(2)}\right] =g\frac{\delta W_{ab}}{\delta
\varphi _{c}}\left( A_{0}^{b}\overline{G}_{c}^{(2)}-\overline{G}%
_{ij}^{(2)b}B_{c}^{ij}+\overline{\gamma }_{c}^{(2)i}H_{i}^{b}\right)  
\nonumber \\
&&+g\frac{\delta ^{2}W_{ab}}{\delta \varphi _{c}\delta \varphi _{d}}\left(
A_{0}^{b}\overline{\gamma }_{c}^{(2)i}B_{d0i}-\frac{1}{2}B_{c}^{ij}\overline{%
\gamma }_{d[i}^{(2)}A_{j]}^{b}\right) ,  \label{xx114}
\end{eqnarray}
\begin{equation}
\left[ H,G_{ij}^{(1)a}\right] =\overline{G}_{ij}^{(2)a},  \label{xx115}
\end{equation}
\begin{equation}
\left[ H,\overline{G}_{ij}^{(2)a}\right] =g\left( \frac{\delta W_{bc}}{%
\delta \varphi _{a}}\overline{G}_{ij}^{(2)c}A_{0}^{b}+\frac{1}{2}\frac{%
\delta ^{2}W_{bc}}{\delta \varphi _{a}\delta \varphi _{d}}A_{0}^{b}\overline{%
\gamma }_{d[i}^{(2)}A_{j]}^{c}\right) ,  \label{xx116}
\end{equation}
\begin{equation}
\left[ H,\gamma _{a}^{(1)i}\right] =\overline{\gamma }_{a}^{(2)i},
\label{xx117}
\end{equation}
\begin{equation}
\left[ H,\overline{\gamma }_{a}^{(2)i}\right] =g\frac{\delta W_{ab}}{\delta
\varphi _{c}}A_{0}^{b}\overline{\gamma }_{c}^{(2)i}.  \label{xx118}
\end{equation}

At this point, we show that the resulting deformations are nontrivial. It is
known that trivial deformations can be eliminated by some field
redefinitions 
\begin{equation}
z^{A}\rightarrow z^{\prime A}=z^{A}+g\lambda ^{A}+O\left( g^{2}\right) ,
\label{xx118a}
\end{equation}
where $\lambda ^{A}$ and the higher-order contributions are in general
nonlinear functions of $z^{A}$. Now, we invoke the requirement of locality,
that plays a key role in quantum field theory. In view of this, we cannot
stress enough that the field redefinitions (\ref{xx118a}) should be local,
because otherwise we cannot transform the deformed theory into the initial
free one, which essentially are both local. Initially, we focus on the
deformations of the constraints. Due to the fact that only the secondary
constraints are modified, we mainly restrict ourselves to this case. Under
the redefinition (\ref{xx118a}), the first-class constraints (\ref{xx94}--%
\ref{xx96}) transform like
\begin{equation}
\overline{G}_{a}^{(2)}\rightarrow \overline{G}_{a}^{\prime (2)}\equiv
-\partial _{i}B_{a}^{0i}-g\partial _{i}\stackrel{(B)}{\lambda }%
_{a}^{i}+gf_{a}\left( z^{A}\right) +O\left( g^{2}\right) \approx 0,
\label{xx118b}
\end{equation}
\begin{equation}
\overline{G}_{ij}^{(2)a}\rightarrow \overline{G}_{ij}^{\prime (2)a}\equiv -%
\frac{1}{2}\partial _{\left[ i\right. }A_{\left. j\right] }^{a}-\frac{g}{2}%
\partial _{\left[ i\right. }\stackrel{(A)}{\lambda }_{\left. j\right]
}^{a}+gf_{ij}^{a}\left( z^{A}\right) +O\left( g^{2}\right) \approx 0,
\label{xx118c}
\end{equation}
\begin{equation}
\overline{\gamma }_{a}^{(2)i}\rightarrow \overline{\gamma }_{a}^{\prime
(2)i}\equiv -\partial ^{i}\varphi _{a}-g\partial ^{i}\stackrel{(\varphi )}{%
\lambda }_{a}+gf_{a}^{i}\left( z^{A}\right) +O\left( g^{2}\right) \approx 0,
\label{xx118d}
\end{equation}
where the functions denoted by $f$ do not contain derivatives of the fields,
and are obtained from (\ref{xx94}--\ref{xx96}) via (\ref{xx118a}) (their
concrete form can be easily written down, but it is not illuminating in this
context). Requiring now that the redefined constraints lead back to those of
the free theory, namely, the latter constraints in (\ref{xx2}--\ref{xx4}),
we find at order one in the coupling constant the equations
\begin{equation}
-\partial _{i}\stackrel{(B)}{\lambda }_{a}^{i}+f_{a}\left( z^{A}\right) =0,
\label{xx118e}
\end{equation}
\begin{equation}
-\frac{1}{2}\partial _{\left[ i\right. }\stackrel{(A)}{\lambda }_{\left.
j\right] }^{a}+f_{ij}^{a}\left( z^{A}\right) =0,  \label{xx118f}
\end{equation}
\begin{equation}
-\partial ^{i}\stackrel{(\varphi )}{\lambda }_{a}+f_{a}^{i}\left(
z^{A}\right) =0.  \label{xx118g}
\end{equation}
As the functions $f_{a}$, $f_{ij}^{a}$ and $f_{a}^{i}$ do not involve the
derivatives of the fields, it follows that the last equations cannot be
satisfied by some local functions $\stackrel{(B)}{\lambda }_{a}^{i}$, $%
\stackrel{(A)}{\lambda }_{i}^{a}$ and $\stackrel{(\varphi )}{\lambda }_{a}$.
This means that we cannot perform a \textit{local} transformation of the
fields that switches the deformed constraints to their initial form, so the
constraints of the interacting theory are indeed nontrivial. Regarding the
reducibility functions, it is simple to see that the deformed functions (\ref
{xx105}--\ref{xx106}) and (\ref{xx110}--\ref{xx111}) reduce to the original
ones (\ref{xx14}--\ref{xx15}) if and only if $W_{ab}\left( \varphi
_{a}\right) =0$. However, the last equations cannot be implied by any field
redefinition, so the reducibility functions of the coupled model are also
nontrivial. Related to the deformed first-class Hamiltonian (\ref{xx112}),
we observe that it is a combination of deformed first-class constraints, so
the result that the modified first-class constraints cannot be brought to
their initial form by a local field redefinition then passes on to the
first-class Hamiltonian. In conclusion, the Hamiltonian deformation of the
free model under study results in a nontrivial gauge theory with an open
(nonlinear) gauge algebra. As the first-class constraints generate gauge
transformations, we expect that the Lagrangian gauge transformations of the
resulting theory are modified with respect to the initial ones.

After some computation, we find that the Lagrangian action of the
interacting theory is expressed by 
\begin{equation}
S[A_{\mu }^{a},H_{\mu }^{a},\varphi _{a},B_{a}^{\mu \nu }]=\int d^{4}x\left(
H_{\mu }^{a}D^{\mu }\varphi _{a}+\frac{1}{2}B_{a}^{\mu \nu }\overline{F}%
_{\mu \nu }^{a}\right) ,  \label{xx119}
\end{equation}
and it is invariant under the gauge transformations 
\begin{equation}
\delta _{\epsilon }A_{\mu }^{a}=\left( D_{\mu }\right) _{\;\;b}^{a}\epsilon
^{b},  \label{xx121}
\end{equation}
\begin{eqnarray}
&&\delta _{\epsilon }H_{\mu }^{a}=\left( D^{\nu }\right)
_{\;\;b}^{a}\epsilon _{\mu \nu }^{b}-g\frac{\delta W_{bc}}{\delta \varphi
_{a}}\epsilon ^{b}H_{\mu }^{c}  \nonumber \\
&&+\frac{g}{2}\frac{\delta ^{2}W_{cd}}{\delta \varphi _{a}\delta \varphi _{b}%
}A^{c\nu }A^{d\rho }\epsilon _{b\mu \nu \rho }+g\frac{\delta ^{2}W_{cd}}{%
\delta \varphi _{a}\delta \varphi _{b}}\epsilon ^{c}A^{d\nu }B_{b\mu \nu },
\label{xx122}
\end{eqnarray}
\begin{equation}
\delta _{\epsilon }\varphi _{a}=-gW_{ab}\epsilon ^{b},  \label{xx120}
\end{equation}
\begin{equation}
\delta _{\epsilon }B_{a}^{\mu \nu }=\left( D_{\rho }\right)
_{a}^{\;\;b}\epsilon _{b}^{\mu \nu \rho }+gW_{ab}\epsilon ^{b\mu \nu }-g%
\frac{\delta W_{ab}}{\delta \varphi _{c}}\epsilon ^{b}B_{c}^{\mu \nu }.
\label{xx123}
\end{equation}
As we have anticipated, the deformation of the initial gauge transformations
(see (\ref{xxgauge})) is essentially due to the deformation of the
first-class constraints. The gauge transformations of the interacting model
are also second-stage reducible, like the original ones, but the
reducibility relations take place on-shell. We observe that neither the
interacting action (\ref{xx119}), nor its gauge transformations, contain the
four-dimensional antisymmetric symbol, as expected. Moreover, although the
gauge structure of the coupled model in four dimensions is richer than that
of the one in two dimensions \cite{mpla}, the Lagrangian in (\ref{xx119})
has the same expression. This emphasises the possibility to construct
nonlinear gauge theories also in dimensions higher than four.

\section{Conclusion}

In conclusion, in this paper we have investigated the consistent Hamiltonian
interactions that can be introduced among a set of scalar fields, two types
of one-forms and a system of two-forms in four dimensions, which are
described in the free limit by an abelian BF theory. Our procedure relies on
the deformation of both BRST charge and BRST-invariant Hamiltonian of the
free version of this model. Related to the deformation of the BRST charge,
we find that only its first-order deformation is non-trivial, while its
consistency reveals the Jacobi identity for a nonlinear algebra. Concerning
the deformation of the BRST-invariant Hamiltonian, we infer again that only
its first-order deformation is non-vanishing. From these two deformed
quantities we derive the first-class constraints, accompanying reducibility
functions, first-class Hamiltonian and modified gauge algebra relations of
the interacting model, which is precisely a four-dimensional nonlinear gauge
theory. This is an example of deformation that modifies the gauge
transformations, reducibility relations, as well as the gauge algebra. This
result generalizes the two-dimensional analysis exposed in \cite{mpla} in
the sense that although the gauge structure of the four-dimensional model is
richer, the Lagrangian of the interacting theory has an expression similar
to that from the two-dimensional case. In this light, there is hope that it
would be possible to use our deformation procedure in order to construct
nonlinear gauge theories in dimensions higher than four.

\section*{Acknowledgment}

This work has been supported by a Romanian Council for Academic Scientific
Research (CNCSIS) grant.


\begin{thebibliography}{99}
\bibitem{2}  M. Henneaux, C. Teitelboim, Quantization of Gauge Systems
(Princeton University Press, Princeton 1992)

\bibitem{8}  E. S. Fradkin, G. A. Vilkovisky, Phys. Lett. \textbf{B55}
(1975) 224; I. A. Batalin, G. A. Vilkovisky, Phys. Lett. \textbf{B69} (1977)
309; E. S. Fradkin, T. E. Fradkina, Phys. Lett. \textbf{B72} (1978) 343; I.
A. Batalin, E. S. Fradkin, Phys. Lett. \textbf{B122} (1983) 157; M.
Henneaux, Phys. Rep. \textbf{126} (1985) 1

\bibitem{9}  R. Ferraro, M. Henneaux, M. Puchin, J. Math. Phys. \textbf{34}
(1993) 2757

\bibitem{10}  G. Barnich, Mod. Phys. Lett. \textbf{A9} (1994) 665

\bibitem{11}  G. Barnich, M. Henneaux, J. Math. Phys. \textbf{37} (1996) 5273

\bibitem{12}  C. Bizdadea, Acta Phys. Polon. \textbf{B32} (2001) 2843;
C. Bizdadea, E. M. Cioroianu, S. O. Saliu, Class. Quantum Grav. 
\textbf{17} (2000) 2007; C. Bizdadea, L. Saliu, S. O. Saliu, 
Int. J. Mod. Phys. \textbf{A15} (2000) 893; C. Bizdadea,
S. O. Saliu, Phys. Scripta \textbf{62} (2000) 261

\bibitem{13}  For a review, see D. Birmingham, M. Blau, M. Rakowski, G.
Thompson, Phys. Rep. \textbf{209} (1991) 129

\bibitem{mpla}  C. Bizdadea, Mod. Phys. Lett. \textbf{A15} (2000) 2047

\bibitem{14}  N. Ikeda, K. I. Izawa, Prog. Theor. Phys. \textbf{89} (1993)
223; 1077; \textbf{90} (1993) 237; N. Ikeda, Ann. Phys. \textbf{235} (1994)
235; P. Schaller, T. Strobl, Phys. Lett. \textbf{B337} (1994) 266;
hep-th/9411163; in Lect. Notes Phys., Vol. \textbf{469} (Springer Verlag,
Berlin 1996) 321; T. Kl\"{o}sch, P. Schaller, T. Strobl, Helv. Phys. Acta 
\textbf{69} (1996) 305; W. Kummer, H. Liebl, D. V. Vassilevich, Nucl. Phys. 
\textbf{B493} (1997) 491

\bibitem{15}  D. J. Gross, A. A. Migdal, Nucl. Phys. \textbf{B340} (1990)
333; T. Banks, M. O'Loughlin, Nucl. Phys. \textbf{B362} (1991) 649; E.
Witten, Proceedings New York 1991--- Differential Geometry Methods in
Theoretical Physics, Vol.\textbf{1} (World Scientific, Singapore 1991) 176;
G. Grignani, G. Nardelli, Class. Quantum Grav. \textbf{10} (1993) 2569; T.
Fujiwara, Y. Igaraschi, J. Kubo, T. Tabei, Phys. Rev. \textbf{D48} (1993)
1736; T. Fujiwara, T. Tabei, Y. Igaraschi, K. Maeda, J. Kubo, Mod. Phys.
Lett. \textbf{A8} (1993) 2147; T. Strobl, Int. J. Mod. Phys. \textbf{D3}
(1994) 281; R. Jackiw, gr-qc/9511048

\bibitem{16}  K. Schoutens, A. Sevrin, P. van Nieuwenhuizen, Commun. Math.
Phys. \textbf{124} (1989) 87; S. Hirano, Y. Kazama, Y. Satoh, Phys. Rev. 
\textbf{D48} (1993) 1687; \"{O}. Dayi, Mod. Phys. Lett. \textbf{A9} (1994)
2157; Int. J. Mod. Phys. \textbf{A12} (1997) 4387; D. Louis-Martinez, G.
Kunstatter, Phys. Rev. \textbf{D52} (1995) 3494; M. Blagojevi\'{c}, M.
Vasili\'{c}, T. Vuka\v {s}inac, Class. Quantum Grav. \textbf{13} (1996)
3003; P. M. Lavrov, P. Y. U. Moshin, Class. Quantum Grav. \textbf{16} (1999)
2247; W. Kummer, G. Tieber, Phys. Rev. \textbf{D59} (1999) 044001; D.
Grumiller, D. Hofmann, W. Kummer, Ann. Phys. \textbf{290} (2001) 69

\bibitem{gen}  G. Barnich, F. Brandt, M. Henneaux, Commun. Math. Phys. 
\textbf{174} (1995) 57; Phys. Rep. \textbf{338} (2000) 439
\end{thebibliography}
\end{document}